\documentclass[12pt]{article}
\usepackage{amsmath}
\usepackage{amssymb}
\usepackage{amsthm}
\usepackage{graphicx}
\usepackage{subfigure}
\usepackage{cite}
\usepackage{url}
\usepackage[small]{caption}
\usepackage{mathtools,xparse}
\theoremstyle{definition}

\begin{document}
\baselineskip 0.6cm

\def\bra#1{\langle #1 |}
\def\ket#1{| #1 \rangle}
\def\inner#1#2{\langle #1 | #2 \rangle}
\def\app#1#2{%
  \mathrel{%
    \setbox0=\hbox{$#1\sim$}%
    \setbox2=\hbox{%
      \rlap{\hbox{$#1\propto$}}%
      \lower1.1\ht0\box0%
    }%
    \raise0.25\ht2\box2%
  }%
}
\def\approxprop{\mathpalette\app\relax}
\DeclarePairedDelimiter{\norm}{\lVert}{\rVert}

\begin{titlepage}

\begin{flushright}
\end{flushright}

\vskip 1.2cm

\begin{center}
{\Large \bf Butterfly Velocities for Holographic Theories of General 
Spacetimes}

\vskip 0.7cm

{\large Yasunori Nomura$^{a,b,c}$ 
and
  Nico Salzetta$^{a,b}$}

\vskip 0.5cm

$^a$ {\it Berkeley Center for Theoretical Physics, Department of Physics,\\
  University of California, Berkeley, CA 94720, USA}

\vskip 0.2cm

$^b$ {\it Theoretical Physics Group, Lawrence Berkeley National Laboratory, 
 CA 94720, USA}

\vskip 0.2cm

$^c$ {\it Kavli Institute for the Physics and Mathematics of the Universe 
 (WPI), University of Tokyo, Kashiwa, Chiba 277-8583, Japan}

\vskip 0.8cm

\abstract{
The butterfly velocity characterizes the spread of correlations in 
a quantum system.  Recent work has provided a method of calculating 
the butterfly velocity of a class of boundary operators using holographic 
duality.  Utilizing this and a presumed extension of the canonical 
holographic correspondence of AdS/CFT, we investigate the butterfly 
velocities of operators with bulk duals living in general spacetimes. 
We analyze some ubiquitous issues in calculating butterfly velocities 
using the bulk effective theory, and then extend the previously proposed 
method to include operators in entanglement shadows.  We explicitly 
compute butterfly velocities for bulk local operators in the holographic 
theory of flat Friedmann-Robertson-Walker spacetimes and find a universal 
scaling behavior for the spread of operators in the boundary theory, 
independent of dimension and fluid components.  This result may suggest 
that a Lifshitz field theory with $z = 4$ is the appropriate holographic 
dual for these spacetimes.}

\end{center}
\end{titlepage}

\section{Introduction}
\label{sec:intro}

The quantum theory of gravity is expected to be formulated in a 
non-gravitational spacetime whose dimension is less than that of the 
bulk gravitational spacetime~\cite{'tHooft:1993gx,Susskind:1994vu,%
Bousso:2002ju}.  The holographic theories for general spacetimes 
are not explicitly known, but we expect that they are strongly coupled 
based on the known holographic correspondence between conformal field 
theories (CFT) and quantum gravity in asymptotically anti-de~Sitter 
(AdS) spacetimes~\cite{Maldacena:1997re}.  If cosmological spacetimes 
do indeed admit holographic descriptions, it is critical to find 
the appropriate dual theories in order to understand the quantum 
nature of gravity in our universe.  In an effort to find such theories, 
we take a bottom-up approach and calculate quantities that can help 
identify them.

A particular quantity that characterizes a strongly coupled system 
is the butterfly velocity~\cite{Shenker:2013pqa,Roberts:2014isa,%
Roberts:2016wdl}, which can be viewed as the effective speed of the 
spread of information relevant for an ensemble of states.  Recently, 
Qi and Yang~\cite{Qi:2017ttv} generalized the concept to general 
subspaces of a Hilbert space, including a code subspace of a 
holographic theory~\cite{Almheiri:2014lwa,Pastawski:2015qua}. 
They then discussed its relationship to the causal structure of 
an emergent bulk theory.

In this paper, we investigate butterfly velocities in holographic 
theories of general spacetimes, described in Refs.~\cite{Nomura:2016ikr,%
Nomura:2017npr}.  In particular, we calculate butterfly velocities 
for bulk local operators in the holographic theory of cosmological 
flat Friedmann-Robertson-Walker (FRW) spacetimes and analyze their 
properties.  We find that they admit a certain universal scaling, 
independent of the fluid component and the dimension of the bulk 
spacetime.  This emerges in the limit that the boundary region 
representing a bulk operator becomes small, where we expect that 
the butterfly velocity reflects properties of the underlying theory.

We also provide an extension of the prescription of 
Ref.~\cite{Qi:2017ttv} for computing the butterfly velocity 
to include more general operators in the bulk.  This generalization 
allows us to calculate the butterfly velocities of bulk local 
operators in some entanglement shadow regions.

Together with the monotonicity property of the change of the volume 
of the holographic space~\cite{Bousso:2015mqa,Bousso:2015qqa,%
Sanches:2016pga} and the behavior of the entanglement entropies of 
subregions of a holographic space~\cite{Nomura:2016ikr,Sanches:2016sxy}, 
our results provide important data for finding explicit holographic 
theories of general spacetimes.  In particular, our results seem 
to indicate a certain relation between spatial and temporal scaling 
in the holographic theory of flat FRW spacetimes.

The organization of the paper is as follows.  In 
Section~\ref{sec:butterfly}, we define the butterfly velocity 
in holographic theories and discuss (extended) prescriptions 
of calculating it using the bulk effective theory.  In 
Section~\ref{sec:FRW}, we compute butterfly velocities in 
the holographic theory of flat FRW universes and analyze their 
properties.  In Section~\ref{sec:discuss}, we discuss possible 
implications of our results.

Throughout the paper, we take units where the bulk Planck length 
is unity.  We assume that the bulk spacetime satisfies the null and 
causal energy conditions.  These impose the conditions $\rho \geq -p$ 
and $|\rho| \geq |p|$, respectively, on the energy density $\rho$ 
and pressure $p$ of an ideal fluid component, so that the equation 
of state parameter, $w = p/\rho$, satisfies $|w| \leq 1$.

\section{Definition of the Butterfly Velocity in Holographic Theories 
of General Spacetimes}
\label{sec:butterfly}

We are interested in the spread of information in holographic theories 
of general spacetimes.  The butterfly velocity is a quantity that 
characterizes the spread of correlations of operators acting within 
a certain subspace of a Hilbert space.  In particular, we can restrict 
our attention to a code subspace of states in which observables 
correspond to operators acting within the bulk effective theory.

We work within the framework described in Ref.~\cite{Nomura:2016ikr}. 
The theory is defined on the holographic spacetime, which for 
a fixed semiclassical bulk spacetime corresponds to a holographic 
screen~\cite{Bousso:1999cb}, a special codimension-1 surface in 
the bulk.  The holographic screen is uniquely foliated by surfaces 
called leaves; this corresponds to a fixed time slicing of the holographic 
theory.  We study how the support of an operator dual to a bulk local 
operator spreads in time.  In Section~\ref{subsec:def}, we follow 
Ref.~\cite{Qi:2017ttv} and define the butterfly velocity in this context. 
We then describe how to calculate it using the bulk effective theory. 
We also discuss conceptual issues associated with this procedure.  In 
Section~\ref{subsec:shadow}, we extend the definition to include bulk 
operators in entanglement shadows.

\subsection{Butterfly velocities on holographic screens}
\label{subsec:def}

We are interested in how the support of a holographic representation 
of a bulk local operator, ${\cal O}$, changes in time in the holographic 
theory.

This analysis is complicated by the fact that each bulk local 
operator can be represented in multiple ways in the holographic 
space (which we may loosely refer to as the boundary, borrowing from 
AdS/CFT language)~\cite{Hamilton:2006az,Czech:2012bh,Dong:2016eik}.  For 
example, suppose the operator is represented over the whole boundary, 
as in the global representation in AdS/CFT.  There is then no concept 
of the operator spreading in time.  Following Ref.~\cite{Qi:2017ttv}, 
we avoid this issue by representing a bulk operator at a point 
$p$ such that $p$ is on the Hubeny-Rangamani-Takayanagi (HRT) 
surface~\cite{Hubeny:2007xt} of a subregion of the boundary. 
Specifically, we consider a subregion $A$ {\it on a leaf} $\sigma_0$ 
(not of an arbitrary spatial section of the holographic screen) 
and represent a bulk local operator ${\cal O}$ located on the HRT 
surface, $\gamma_A$, of $A$.  Based on intuition arising from analyzing 
tensor network models~\cite{Pastawski:2015qua,Hayden:2016cfa}, we 
expect that such a representation is unique.  We denote the operator 
in the boundary theory represented in this way on $A$ as ${\cal O}_A$.

We want to know the spatial region $B$ on the leaf $\sigma_{\varDelta t}$, 
which is in the future of $\sigma_0$ by time $\varDelta t$, such that 
every operator ${\cal B}$ supported on $B$ satisfies
\begin{equation}
  \bra{\Psi_i} [{\cal O}_A, {\cal B}] \ket{\Psi_j} = 0.
\label{eq:comm}
\end{equation}
Here, $\ket{\Psi_i}$ and $\ket{\Psi_j}$ are arbitrary states in the 
code subspace.  Recall that there is a natural way of relating regions 
on different leaves of a holographic screen~\cite{Sanches:2016pga}. 
The spatial coordinates on $\sigma_{\varDelta t}$ can be defined from 
those on $\sigma_0$ by following the integral curves of a vector field 
orthogonal to every leaf on the holographic screen.  We can then define 
the region $A'$ on $\sigma_{\varDelta t}$ corresponding to $A$ on 
$\sigma_0$ by following such curves.  This allows us to define the 
distance, $\varDelta d$, of the operator spread for each point $q$ 
on the boundary, $\partial A'$, of $A'$ as the distance from $q$ to 
the region $B$ in the direction orthogonal to $\partial A'$.

For an arbitrary operator in $A$, there is no reason that the distance 
$\varDelta d$ is independent of the location on $\partial A'$.  The 
butterfly velocity can then be defined using the largest of $\varDelta d$ 
along $\partial A'$~\cite{Qi:2017ttv}:
\begin{equation}
  v_B \equiv \underset{\theta_i}{\rm max} \frac{\varDelta d}{\varDelta t},
\label{eq:vB-def}
\end{equation}
where $\{ \theta_i \}$ are the coordinates of $q$ on $\partial A'$. 

We now discuss how to calculate $v_B$ using the bulk effective theory. 
For this, we must understand how time evolved operators in the bulk 
are represented in the boundary theory.  More specifically, given 
a particular representation, ${\cal O}_A$, of ${\cal O}$ in the 
holographic Hilbert space, what representation of the time evolved 
bulk operator (within the light cone of $p$) does the time evolution 
of ${\cal O}_A$ corresponds to?  Without an explicit boundary theory 
it is not possible to answer this question, but we can still make 
some headway using intuition.  First, we may expect that the region 
$B$ on $\sigma_{\varDelta t}$ (defined above) fully excludes $A'$. 
This is the statement that the support of the operator does not shrink 
in any direction.  Second, we want the ``minimal necessary extension'' 
of the leaf subregion $A$.  For instance, it seems unphysical that 
a bulk operator represented on subregion $A$ should immediately 
time evolve into the full boundary representation of the future bulk 
operator.  We thus seek the correspondingly ``maximal'' region $B$ 
whose entanglement wedge does not contain the interior of light cone 
of $p$.  Excluding the light cone ensures that no information can 
be sent in the bulk which would compromise the commutativity between 
${\cal O}_A$ and ${\cal B}$ within the code subspace.

From these considerations, we come up with two possible procedures for 
calculating the butterfly velocity of a bulk operator at a point $p$:
\begin{enumerate}
\item Maximize the volume of subregion $B'$ subject to the constraint 
that $A' \cap B' = \emptyset$ and that the entanglement wedge of $B'$ does 
not contain the interior of the light cone of $p$.  The resulting 
subregion then gives $B$.
\item Find the subregion $B$ with the distance from $\partial B$ to 
$\partial A'$ being both minimal and independent of the location on 
$\partial B$, again subject to the constraint that the entanglement 
wedge of $B$ does not contain the light cone of $p$.
\end{enumerate}
One can certainly consider other possibilities as well, but these are 
the two most intuitively obvious candidates.  However, we find that 
the first possibility leads to discontinuous behavior of $B$ as $p$ 
moves across the tip of the HRT surface of a spherical cap region. 
We therefore focus on the second possibility, which aligns with 
Ref.~\cite{Qi:2017ttv}.%
\footnote{It is possible that the validity of these procedures may 
 be analyzed by explicitly calculating the boundary dual of bulk 
 local operators located on an HRT surface by using recently proposed 
 methods of entanglement wedge reconstruction~\cite{Faulkner:2017vdd,%
 Cotler:2017erl}.}

Essentially, this possibility postulates that the support of the operator 
${\cal O}_A$ spreads uniformly:
\begin{equation}
  \frac{\partial \varDelta d}{\partial \theta_i} = 0.
\label{eq:uniform}
\end{equation}
We assume that this is indeed the case.  The prescription of calculating 
the butterfly velocity can then be given explicitly as follows.  We 
first consider a region $B'(\varDelta \lambda)$ on $\sigma_{\varDelta t}$ 
which is (i) $\varDelta \lambda$ away from $A'$, i.e.\ the distance from 
any point on $\partial A'$ to $B'(\varDelta \lambda)$ is $\varDelta \lambda$ 
in the direction orthogonal to $\partial A'$ and (ii) the entanglement 
wedge of $B'(\varDelta \lambda)$ does not contain the interior of the 
light cone of $p$.  The butterfly velocity of ${\cal O}_A$ is then obtained 
by finding $B'(\varDelta \lambda)$ with the smallest $\varDelta \lambda$
\begin{equation}
  v_B = \underset{\varDelta \lambda}{\rm min} 
    \frac{\varDelta \lambda}{\varDelta t}.
\label{eq:vB-holo}
\end{equation}
Note that the resulting $v_B$ depends on how the bulk operator ${\cal O}$ 
is represented initially, i.e.\ $A$ and the location of ${\cal O}$ on 
$\gamma_A$.

If the assumption of Eq.~(\ref{eq:uniform}) is not valid in general, 
then our results for the ``off-center'' operators, $f \neq 0$, 
in Section~\ref{sec:FRW} (as well as any related results in 
Ref.~\cite{Qi:2017ttv}) would have to be reinterpreted as representing 
something other than $v_B$ defined in Eq.~(\ref{eq:vB-def}).  However, 
our results for the operators at the tip of the HRT surface, $f = 0$, 
are still correct in this case, since Eq.~(\ref{eq:uniform}) is 
guaranteed by the symmetry of the setup.

\subsection{Bulk operators in entanglement shadows}
\label{subsec:shadow}

In the prescription given in the previous subsection, the bulk operator 
${\cal O}$ was on the HRT surface of a subregion $A$ on a leaf. 
Motivated by the idea that a bulk local operator can be represented in 
the holographic theory not only at an intersection of HRT surfaces but 
also at an intersection of the edge of the entanglement wedges (associated 
with subregions of leaves)~\cite{Nomura:2017npr,Sanches:2017xhn}, we 
expect that we can similarly calculate the butterfly velocity for an 
operator ${\cal O}_A$ corresponding to a bulk operator at a point $p$ 
on the boundary of the entanglement wedge of $A$, ${\rm EW}(A)$.

There is no obstacle in using either of the prescriptions detailed 
in the previous subsection, except now we take $p$ to be on the edge 
of the entanglement wedge.  In this case, we must be careful to exclude 
the {\it entire} light cone of $p$ when finding $B$.  We find that the 
behavior of $v_B$ is qualitatively different depending on whether $p$ 
is on the future or past boundary of ${\rm EW}(A)$.

Suppose $p$ is on the past boundary of ${\rm EW}(A)$.  In this case, 
${\rm EW}(B)$ is not limited by excluding the part of $p$'s light 
cone infinitesimally close to $p$ (as is the case when $p$ is on 
$\gamma_A$), but by the part of the light cone that is just to the 
future of $\gamma_A$.  Aside from this, there is no other new aspect 
compared with the case in which $p$ is on $\gamma_A$.  In particular, 
$A'$ is forced to spread relative to $A$ in both prescriptions.

There is, however, a subtlety when $p$ is on the future boundary 
of ${\rm EW}(A)$.  This arises because ${\rm EW}(\bar{A'})$ automatically 
excludes the light cone of points located on the future boundary 
of ${\rm EW}(A)$.  Here, $\bar{A'}$ is the complement of $A'$ on 
$\sigma_{\varDelta t}$.  A direct application of the first prescription 
from the previous subsection would then result in a butterfly velocity 
of $0$ for bulk operators at all points on the future boundary of 
${\rm EW}(A)$.  This is due to the constraint that $A' \cap B = 
\emptyset$, forcing $v_B \ge 0$.  This constant $v_B = 0$ behavior 
may encourage us to abandon the constraint, but doing so leads to 
severely discontinuous behavior of $B$.  Namely, the resulting region 
$B$ on $\sigma_{\varDelta t}$ is independent of the original region $A$ 
on $\sigma_0$, because $B$ will always find the same global maximum.

The second prescription has more interesting behavior so long as 
we allow for the distance from $\partial B$ to $\partial A'$ to be 
negative.  Doing so, we see that $B$ is now constrained by excluding 
the {\it past} light cone of $p$, and for the resulting $B$, $A' \cap 
B \neq \emptyset$, so that $v_B \le 0$.  This is interesting because 
as we move forward in the boundary time, we are actually tracking the 
past time evolution of a bulk local operator.  The shrinking support 
of ${\cal O}_A$ could indicate that this is a finely tuned boundary 
operator.

This generalization to the boundary of entanglement wedges allows us 
to calculate the butterfly velocity of operator ${\cal O}_A$ representing 
a bulk local operator in an entanglement shadow, i.e.\ a spacetime 
region in which HRT surfaces do not probe.

\section{Butterfly Velocities for the Holographic Theory of FRW Universes}
\label{sec:FRW}

In this section, we compute butterfly velocities for the holographic 
theory of $(3+1)$-dimensional flat Friedmann-Robertson-Walker (FRW) 
universes:
\begin{equation}
  ds^2 = a^2(\eta) \bigl[ -d\eta^2 + dr^2 
    + r^2 (d\psi^2 + \sin^2\!\psi\, d\phi^2) \bigr],
\label{eq:FRW}
\end{equation}
where $a(\eta)$ is the scale factor with $\eta$ being the conformal 
time.  We mainly focus on the case in which a universe is dominated 
by a single ideal fluid component with the equation of state parameter 
$w = p/\rho$ with $|w| \leq 1$.

In Section~\ref{subsec:analytic}, we derive an analytic expression 
for the butterfly velocity of a bulk local operator near the holographic 
screen.  In Section~\ref{subsec:numerical}, we numerically calculate 
the butterfly velocity for a bulk operator located at the tip of an 
HRT surface with an arbitrary depth.  In Section~\ref{subsec:arbitrary}, 
we extend the result of Section~\ref{subsec:analytic} to arbitrary 
spacetime dimensions.

\subsection{Local operators near the holographic screen}
\label{subsec:analytic}

Consider a spherical cap region
\begin{equation}
  \Gamma: 0 \leq \psi \leq \gamma,
\label{eq:s-cap}
\end{equation}
on the leaf at a time $\eta_*$, which is located at
\begin{equation}
  r = \frac{a(\eta)}{\dot{a}(\eta)}\biggr|_{\eta = \eta_*} \equiv r_*.
\label{eq:app-hor}
\end{equation}
Following Ref.~\cite{Nomura:2016ikr}, we go to cylindrical coordinates:
\begin{equation}
  \xi = r \sin\psi,
\qquad
  z = r \cos\psi - r_* \cos\gamma,
\label{eq:cylind}
\end{equation}
in which the boundary of $\Gamma$, $\partial \Gamma$, is located at
\begin{equation}
  \xi = r_* \sin\gamma \equiv \xi_*,
\qquad
  z = 0.
\label{eq:leaf-cyl}
\end{equation}

In the case that $\gamma \ll 1$, i.e.\ $\xi_* \ll r_*$, the HRT surface 
anchored to $\partial \Gamma$ can be expressed in a power series form. 
Denoting the surface by $\eta$ and $z$ as functions of $\xi$, we find
\begin{align}
  \eta(\xi) =& \eta_* + \eta^{(2)}(\xi) + \eta^{(4)}(\xi) + \cdots,
\label{eq:HRT-exp}\\
  z(\xi) =& 0,
\label{eq:z=0}
\end{align}
where
\begin{align}
  \eta^{(2)}(\xi) =& \frac{\dot{a}}{2a} (\xi_*^2 - \xi^2),
\label{eq:exp-2}\\
  \eta^{(4)}(\xi) =& -\frac{\dot{a}}{16a^3} (\xi_*^2 - \xi^2) 
    \bigl\{ 4\dot{a}^2 \xi_*^2 - a\ddot{a}(3\xi_*^2 - \xi^2) \bigr\},
\label{eq:exp-4}
\end{align}
with
\begin{equation}
  a \equiv a(\eta_*),
\qquad
  \dot{a} \equiv \frac{da(\eta)}{d\eta}\biggr|_{\eta = \eta_*},
\qquad
  \ddot{a} \equiv \frac{d^2a(\eta)}{d\eta^2}\biggr|_{\eta = \eta_*}.
\label{eq:a-da-dda}
\end{equation}
We consider a bulk local operator on this surface.

\begin{figure}[t]
\begin{center}
  \includegraphics[height=6cm]{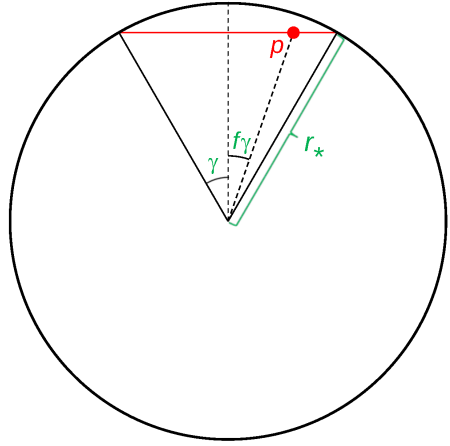}
\end{center}
\caption{A local operator at $p$, represented by the dot, is on the HRT 
 surface anchored to the boundary of a spherical cap region, $0 \leq \psi 
 \leq \gamma$, on the leaf at time $\eta_*$, located at $r = r_*$.  Note 
 that the figure suppresses the time direction; for example, the operator 
 is not at the same time as the leaf.}
\label{fig:f-gamma}
\end{figure}

We parameterize the location, $p$, of the operator by a single number 
$f$ ($0 \leq f < 1$) representing how much fractionally the operator 
is ``off the center,'' i.e.\ the operator is located on the surface
\begin{equation}
  \psi = f \gamma,
\label{eq:op-loc}
\end{equation}
with $\eta$ and $r$ determined by the condition that it is also on the 
HRT surface of Eq.~(\ref{eq:HRT-exp}); see Fig.~\ref{fig:f-gamma}.
(The value of $\phi$ is arbitrary because of the symmetry of the problem; 
below we take $\phi = 0$ without loss of generality.)  In cylindrical 
coordinates, this implies that the location of the operator, $(\eta, 
\xi, z) = (\eta_B, \xi_B, z_B)$, is given by
\begin{align}
  \eta_B =& \eta_* + \frac{\dot{a}}{2a} (\xi_*^2 - \xi_B^2) 
    - \frac{\dot{a}}{16a^3} (\xi_*^2 - \xi_B^2) 
    \bigl\{ 4\dot{a}^2 \xi_*^2 - a\ddot{a}(3\xi_*^2 - \xi_B^2) \bigr\},
\label{eq:eta_B}\\
  \xi_B =& \frac{\tan(f\gamma)}{\tan\gamma} \xi_*,
\label{eq:xi_B}\\
  z_B =& 0,
\label{eq:z_B}
\end{align}
where we have ignored the terms higher order than $\eta^{(4)}(\xi)$ 
in Eq.~(\ref{eq:HRT-exp}), which are not relevant for our leading order 
calculation.  The future light cone associated with $p$ is then 
given by
\begin{equation}
  \eta = \eta_B + \sqrt{(x-\xi_B)^2 + y^2 + z^2},
\label{eq:light-cone}
\end{equation}
where we have introduced the coordinates $x = \xi \cos\phi$ and $y = \xi 
\sin\phi$.

In order to derive the butterfly velocity for the operator at $p$, we 
need to find the smallest spherical cap region on the leaf at $\eta = 
\eta_* + \delta\eta$
\begin{equation}
  \Gamma': 0 \leq \psi \leq \gamma + \delta\gamma,
\label{eq:s-cap'}
\end{equation}
so that the entanglement wedge associated with the complement of 
$\Gamma'$ on the leaf does not contain the interior of the future 
light cone of $p$, Eq.~(\ref{eq:light-cone}).  This occurs for 
the value of $\delta\gamma$ at which the HRT surface anchored to 
$\partial \Gamma'$ 
\begin{align}
  \eta(\xi) =& \eta_* + \frac{\dot{a}}{2a} (\xi_*^2 - \xi^2) 
    - \frac{\dot{a}}{16a^3} (\xi_*^2 - \xi^2) 
    \bigl\{ 4\dot{a}^2 \xi_*^2 - a\ddot{a}(3\xi_*^2 - \xi^2) \bigr\} 
\nonumber\\
  & {} + \delta\eta 
    - \frac{\dot{a}^2}{2a^2} 
      \biggl( 1 - \frac{a\ddot{a}}{\dot{a}^2} \biggr) 
      (\xi_*^2 - \xi^2) \delta\eta 
\nonumber\\
%  & \quad {} + \frac{\dot{a}^2}{16a^4} (\xi_*^2 - \xi^2) 
%      \biggl\{ 12\dot{a}^2 \xi_*^2 - 2a\ddot{a} (7\xi_*^2-\xi^2) 
%        + \frac{a^2\dddot{a}}{\dot{a}} (3\xi_*^2-\xi^2) \biggr\} \delta\eta 
%\nonumber\\
  & {} + \frac{\dot{a}}{a} \xi_* \delta\xi_* 
    - \frac{\dot{a}}{4a^3} \bigl\{ 2\dot{a}^2 (2\xi_*^2 - \xi^2) 
      - a\ddot{a} (3\xi_*^2 - 2\xi^2) \bigr\} \xi_* \delta\xi_*,
\label{eq:HRT'-eta}\\
  z(\xi) =& \biggl( 1 - \frac{a\ddot{a}}{\dot{a}^2} \biggr) 
      \cos\gamma\, \delta\eta
    - \frac{a}{\dot{a}} \sin\gamma\, \delta\gamma,
\label{eq:HRT'-z}
\end{align}
is tangent to the light cone.  Here,
\begin{equation}
  \delta\xi_* = \biggl( 1 - \frac{a\ddot{a}}{\dot{a}^2} \biggr) 
      \sin\gamma\, \delta\eta
    + \frac{a}{\dot{a}} \cos\gamma\, \delta\gamma,
\label{eq:delta-xi*}
\end{equation}
and we have suppressed (some of) the terms that do not contribute to the 
leading order result.

The conditions for the tangency are given by%
\footnote{We would like to thank Yiming Chen, Xiao-Liang Qi, and 
 Zhao Yang for correcting the wrong tangency condition in a previous 
 version.  The results now agree with the monotonicity statement in 
 Ref.~\cite{Qi:2017ttv}, which we believed did not apply to our setup.}
\begin{align}
  \eta(x) &= \eta_B + \sqrt{(x-\xi_B)^2 + y^2 + z(x)^2},
\label{eq:tang-1}\\
  \frac{d\eta(x)}{dx} &= \frac{x-\xi_B}{\sqrt{(x-\xi_B)^2 + y^2 + z(x)^2}},
\label{eq:tang-2}\\
  y &= 0,
\label{eq:tang-3}
\end{align}
where the functions $\eta(x)$ and $z(x)$ are given by 
Eqs.~(\ref{eq:HRT'-eta}) and (\ref{eq:HRT'-z}).  These yield the 
relation between $\delta\eta$ and $\delta\gamma$
\begin{equation}
  \delta\eta 
  = \frac{\dot{a}^2}{4a^2} (3\xi_*^2 - 2\xi_B^2) \xi_* \delta\gamma,
\label{eq:delta-rel}
\end{equation}
as well as the location in which the HRT surface touches the light cone
\begin{equation}
  x = \xi_B - \frac{\dot{a}}{a} \xi_* \xi_B \delta\gamma.
\label{eq:touch}
\end{equation}
Using Eq.~(\ref{eq:xi_B}), Eq.~(\ref{eq:delta-rel}) becomes
\begin{equation}
  \frac{\delta\gamma}{\delta\eta} 
  = \frac{4\dot{a}}{a} \frac{1}{3-2f^2} \frac{1}{\gamma^3},
\label{eq:vel-1}
\end{equation}
where we have used $\xi_* = \gamma a/\dot{a}$; see Eqs.~(\ref{eq:app-hor}) 
and (\ref{eq:leaf-cyl}).  Representing the butterfly velocity $v_B$ 
in terms of the coordinate distance along the holographic space, 
$\delta \lambda = r_* \delta\gamma$, and the conformal time, we finally 
obtain
\begin{equation}
  v_B \equiv \frac{\delta \lambda}{\delta\eta} 
  = \frac{4}{3-2f^2} \frac{1}{\gamma^3}.
\label{eq:vel-pert}
\end{equation}

\begin{figure}[t]
\begin{center}
  \includegraphics[height=6.5cm]{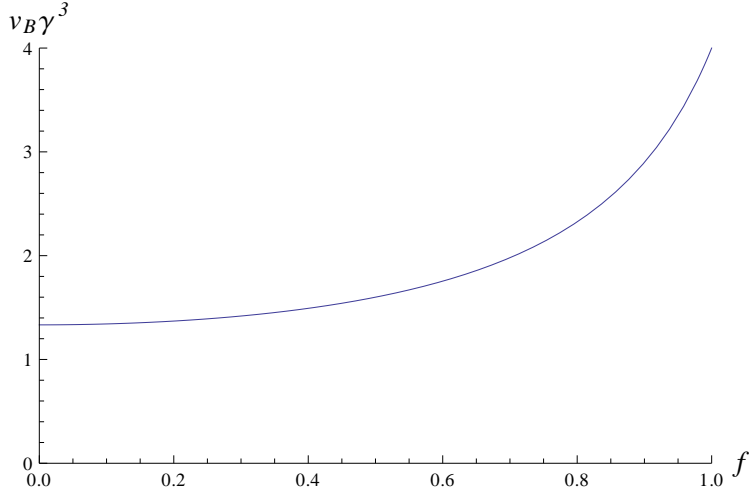}
\end{center}
\caption{Butterfly velocity $v_B$ multiplied by $\gamma^3$ as a function 
 of $f$.  Here, $\gamma$ is the angular size of the leaf region, and 
 $f$ is the fractional displacement of the bulk local operator from 
 the tip of the HRT surface; see Eq.~(\ref{eq:op-loc}).}
\label{fig:v_B}
\end{figure}

There are several features one can see in Eq.~(\ref{eq:vel-pert}). 
First, the butterfly velocity is non-negative, $v_B \geq 0$, as expected. 
Second, for $\gamma \ll 1$, which we are focusing on here, the butterfly 
velocity is much faster than the speed of light.  In fact, it diverges 
as $\gamma \rightarrow 0$ with the specific power of $\gamma^{-3}$. 
When the operator is at the tip of the HRT surface, i.e.\ $f = 0$, the 
butterfly velocity takes the particularly simple form
\begin{equation}
  v_B|_{f=0} = \frac{4}{3} \frac{1}{\gamma^3}.
\label{eq:f=0}
\end{equation}
In Fig.~\ref{fig:v_B}, we plot $v_B \gamma^3$ as a function of $f$. 
We find that the butterfly velocity increases as the operator moves 
closer to the holographic screen:
\begin{equation}
  \frac{dv_B}{df} 
  = \frac{16f}{(3-2f^2)^2} \frac{1}{\gamma^3} 
  > 0.
\label{eq:dv_B-df}
\end{equation}
This is consistent with the monotonicity result in Ref.~\cite{Qi:2017ttv}.

It is interesting that the scale factor has completely dropped out from 
the final expression of Eq.~(\ref{eq:vel-pert}).  This implies that 
regardless of the content of the universe, the short distance behavior 
of the butterfly velocity is universal in the holographic theory 
of flat FRW spacetimes.  As we will see in the next subsection, the 
butterfly velocity's dependence on the scale factor appears as we move 
away from the $\gamma \ll 1$ limit.  This suggests that the details 
of the FRW bulk physics are related with long distance effects in the 
holographic theory.

\subsection{Local operators at arbitrary depths}
\label{subsec:numerical}

Beyond the $\gamma \ll 1$ limit, we must resort to a numerical method 
in order to solve for the butterfly velocity.  For this purpose, we 
focus on the case in which the universe is dominated by a single ideal 
fluid component with the equation of state parameter $w$.  In this case, 
the scale factor behaves as
\begin{equation}
  a(\eta) \propto 
    \left\{ \begin{array}{l}
      \eta^{\frac{2}{1+3 w}} \\
      e^{c \eta}\,\, (c>0)
    \end{array} \right.
\quad\mbox{for}\quad
  \eta \,\left\{\! \begin{array}{l} 
    \neq -\frac{1}{3} \\ = -\frac{1}{3} \end{array}. \right.
\label{eq:a-eta}
\end{equation}
When the universe is dominated by a single fluid component, the 
butterfly velocity $v_B$, expressed in terms of angle $\gamma$, does 
not depend on time $\eta$.  This can be seen by using appropriate 
coordinate transformations, in a way analogous to the argument in 
Section~III\,A\,1 of Ref.~\cite{Nomura:2016ikr} showing that a screen 
entanglement entropy normalized by the leaf area does not depend 
on time.

\begin{figure}[t]
\begin{center}
  \includegraphics[height=6.5cm]{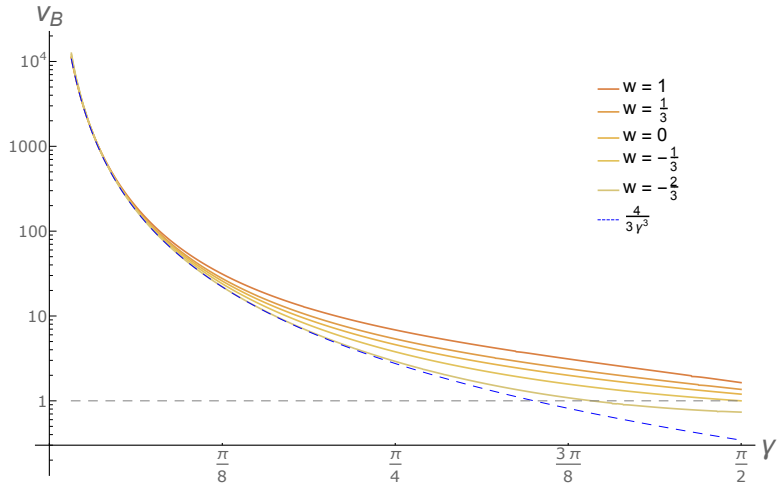}
\end{center}
\caption{Butterfly velocity $v_B$ of a bulk local operator at the tip 
 of the HRT surface, $f=0$, as a function of the angular size $\gamma$ 
 of the leaf region for $w = 1$, $1/3$, $0$, $-1/3$, and $-2/3$ (solid 
 curves, from top to bottom).  The dashed curve represents $v_B = 
 4/3\gamma^3$, the analytic result obtained for $\gamma \ll 1$ in 
 Eq.~(\ref{eq:f=0}).  The horizontal dashed line represents the speed 
 of light.}
\label{fig:v_B_f=0}
\end{figure}
In Fig.~\ref{fig:v_B_f=0}, we show the results of our numerical 
calculations of the butterfly velocity, $v_B$, as a function of $\gamma$ 
for a bulk operator located on the tip of the HRT surface, $f=0$, for 
$w = 1$, $1/3$, $0$, $-1/3$, and $-2/3$.  We find that beyond $\gamma 
\ll 1$, the butterfly velocity deviates from the limiting expression 
of Eq.~(\ref{eq:f=0}), which is depicted by the dashed curve.  In fact, 
the functional form of $v_B|_{f=0}(\gamma)$ is not universal and depends 
on $w$.

We find that for sufficiently large values of $w$ the butterfly velocity 
$v_B|_{f=0}$ is always faster than the speed of light (depicted by 
the horizontal dashed line), while for smaller values of $w$ it can 
be slower than the speed of light for $\gamma$ close to $\pi/2$ (i.e.\ 
when the subregion on the leaf becomes large, approaching a hemisphere). 
The boundary between the two behaviors lies at $w = -1/3$, when the 
expansion of the universe changes between deceleration and acceleration.

\subsection{Arbitrary spacetime dimensions}
\label{subsec:arbitrary}

There is no obstacle in performing the same calculations as in the previous 
subsections in arbitrary spacetime dimensions.  Here we present the 
analytic results corresponding to those in Section~\ref{subsec:analytic} 
for $(d+1)$-dimensional flat FRW universes.

The butterfly velocity, corresponding to Eq.~(\ref{eq:vel-pert}), is 
given by
\begin{equation}
  v_B = \frac{2}{\frac{d+3}{d+1}-f^2} \frac{1}{\gamma^3}.
\label{eq:vel-pert_gen-dim}
\end{equation}
Again, this is non-negative and does not depend on the scale factor. 
We also find that the exponent of $\gamma$ is universal
\begin{equation}
  v_B \sim \frac{1}{\gamma^3},
\label{eq:scaling}
\end{equation}
regardless of the spacetime dimension.  The $f$ dependence of $v_B$ 
is given by
\begin{equation}
  \frac{dv_B}{df} 
  = \frac{4f}{\bigl( \frac{d+3}{d+1}-f^2 \bigr)^2} 
    \frac{1}{\gamma^3} 
  > 0,
\label{eq:dv_B-df_gen-dim}
\end{equation}
which is consistent with the monotonicity result of Ref.~\cite{Qi:2017ttv}.

\section{Discussion}
\label{sec:discuss}

Our investigation has used a definition of butterfly velocity that 
differs from that in the literature regarding lattice systems and 
spin chains.  The main difference is that in our case, the excitations 
of concern (in the boundary theory) are not local operators.  They 
have support on a large subregion of the space.  This is in contrast 
to the lattice definition which considers commutators of local operators 
separated in space and time.  But the conceptual overlap is clear; we 
are concerned with when and where operators commute.  The investigation 
of this paper allows us to find the effective ``light cone'' in the 
holographic theory.

Sending $\gamma \rightarrow 0$ would correspond to a local operator 
in the holographic theory, and the result that the butterfly velocity 
diverges in this limit may seem to indicate that the holographic theory 
is highly nonlocal.  However, this is not necessarily the case, 
as the divergent velocity is integrable.  By setting $f = 0$ in 
Eq.~(\ref{eq:vel-pert_gen-dim}) and converting $\lambda$ to 
$\gamma$ (see, e.g., Eqs.~(\ref{eq:vel-1}) and (\ref{eq:vel-pert})), 
we obtain
\begin{equation}
  \frac{d\gamma}{d\eta} 
  = \frac{q_{d}}{\eta} \frac{2(d+1)}{(d+3)\gamma^3},
\label{eq:velocity-angle}
\end{equation}
where $q_d = 2/(d-2+d w)$.  From this expression, we find
\begin{equation}
  \gamma(t_{\rm H}) 
  = \left[ \frac{8(d+1)}{d+3} t_{\rm H} \right]^{\frac{1}{4}},
\label{eq:growth}
\end{equation}
where $t_{\rm H} = q_d \ln(\eta/\eta_i)$, the number of Hubble times 
elapsed since the excitation. This shows that the light cone spreads 
like $t^{1/4}$, regardless of dimension.

Sub-linear growth like this is not an uncommon phenomenon in physics. 
A localized heat source subject to the heat equation will diffuse 
as $t^{1/2}$.  Even spin chain systems where the Lieb-Robinson bound 
applies (and suggests a linear dispersion) can admit power law 
behavior for the effective growth of operators~\cite{Luitz:2017jrn}. 
The specific relationship of $\varDelta x \sim \varDelta t^{1/4}$ 
suggests that we should be looking for a theory with dynamical 
exponent $z = 4$, and the fact that this holds regardless of spacetime 
dimension may indicate that a Lifshitz field theory with $z = 4$ is 
the appropriate dual theory for flat FRW spacetimes.  Note that results 
from Ref.~\cite{Qi:2017ttv} show that $v_B \rightarrow 1$ as $\gamma 
\rightarrow 0$ for asymptotically AdS spacetimes.  Similarly analyzing 
this result would suggest that a $z = 1$ theory is the appropriate 
dual for AdS, as is indeed the case. These ideas will be investigated 
in future work.

\section*{Acknowledgments}

We thank Yiming Chen, Xiao-Liang Qi, and Zhao Yang for identifying 
an error in a previous version.  We also thank Snir Gazit, Bryce Kobrin, 
Johannes Motruk, Dan Parker, Pratik Rath, Fabio Sanches, Romain Vasseur, 
and Aron Wall for useful discussions.  N.S. thanks Kavli Institute for 
the Physics and Mathematics of the Universe, University of Tokyo for 
hospitality during the visit in which a part of this work was carried 
out.  This work was supported in part by the National Science Foundation 
under grant PHY-1521446, by the Department of Energy, Office of Science, 
Office of High Energy Physics under contract No.\ DE-AC02-05CH11231, 
and by MEXT KAKENHI Grant Number 15H05895.

\end{document}